\pdfoutput=1
\documentclass[sigconf]{acmart}

\AtBeginDocument{%
  }

\copyrightyear{2025}
\acmYear{2025}
\setcctype{by-nc}
\acmConference[CHI '25]{CHI Conference on Human Factors in Computing Systems}{April 26-May 1, 2025}{Yokohama, Japan}
\acmBooktitle{CHI Conference on Human Factors in Computing Systems (CHI '25), April 26-May 1, 2025, Yokohama, Japan}\acmDOI{10.1145/3706598.3713666}
\acmISBN{979-8-4007-1394-1/25/04}

\begin{document}

\title{"They've Over-Emphasized That One Search": Controlling Unwanted Content on TikTok's For You Page}

\author{Julie A. Vera}
\authornote{Both authors contributed equally to this research.}
\email{jvera@uw.edu}
\orcid{0000-0002-4154-9273}
\affiliation{%
  \institution{Department of Human Centered Design and Engineering, University of Washington}
  \city{Seattle}
  \state{Washington}
  \country{USA}
}

\author{Sourojit Ghosh}
\authornotemark[1]
\email{ghosh100@uw.edu}
\orcid{0000-0001-5143-6187}
\affiliation{%
  \institution{Department of Human Centered Design and Engineering, University of Washington}
  \city{Seattle}
  \state{Washington}
  \country{USA}
}

\renewcommand{\shortauthors}{Vera J.A. et al.}

\begin{abstract}
Modern algorithmic recommendation systems seek to engage users through behavioral content-interest matching. While many platforms recommend content
based on engagement metrics, others like TikTok deliver interest-based content, resulting in recommendations perceived to be hyper-personalized compared to other platforms. TikTok's robust recommendation engine has led some users
to suspect that the algorithm knows users “better than they know themselves," but this is not always true. In this paper, we explore TikTok users’ perceptions of recommended content on their For You Page (FYP), specifically
calling attention to unwanted recommendations. Through qualitative interviews of 14 current and former TikTok users, we find themes of frustration with recommended content, attempts to rid themselves of unwanted content, and various degrees of success in eschewing such content. We discuss implications in the larger context of folk theorization and contribute concrete tactical and behavioral examples of \textit{algorithmic persistence}.
\end{abstract}

\ccsdesc[500]{Human-centered computing~Empirical studies in collaborative and social computing}
\ccsdesc[500]{Human-centered computing~Collaborative filtering}
\ccsdesc[500]{Human-centered computing~Social media}

\keywords{algorithms, TikTok, folk theories, content recommendation, social media}

\maketitle

\section{Introduction}
Social media users are now acutely aware of the recommendation engines that allow platforms to serve hyper-personalized content.
TikTok's logic is centered around its For You Page (FYP), the primary feature through which users interact with content. The FYP's content-interest matching has led some users to theorize that the TikTok algorithm “knows you better than you know yourself” \cite{eslami2016first}. As users co-create algorithmic identities \cite{cheney2011new} through negotiations with recommended content, there will inevitably be some pushback against recommendations that do not resonate. 

In this exploratory study with 14 first-hand accounts of current/former TikTok users, we document theories of acquiring  \textit{unwanted} content and tactics employed to remove such content. We explore the following research questions:

\begin{itemize}
\item RQ1: What are some folk theories associated with encountering difficult-to-remove FYP content?
\item RQ2: What are the tactics that TikTok users employ to remove unwanted content from their For You Pages?
\end{itemize}

We uncover experiences on TikTok where users not only receive content that is poorly matched to their interests but also situations where they are not able to prevent their FYPs from recommending genres of content in which they have, seemingly, explicitly signaled negative interest. We discuss these findings in the context of folk theorization and present evidence for a phenomenon on TikTok we call \textit{algorithmic persistence}: where users have little to no success in removing unwanted content from their FYPs despite their best efforts. 

\section{Related Work}
{\subsection{Folk Theories and Folk Theorization}
Folk theories regarding social media and algorithms are best described as collectively/individually-observed ideas or \textit{theories} of the way social media platforms \cite{eslami2016first,young2023young}, algorithms \cite{devito2021adaptive, dogruel2021folk, karizat2021algorithmic, simpson2022tame}, or moderation practices \cite{mayworm2024content, moran2022folk} work. Such theories are not based on definitive evidence from platform designers themselves but are credible and seemingly plausible because they are grounded in the lived experiences of everyday users \cite{devito2017algorithms, devito2018people, french2017s, ytre2021folk}.}

Previous research has specifically described the experiences of both average-audience \cite{french2017s, ytre2021folk} and marginalized users \cite{devito2018people, devito2022transfeminine, mayworm2024content, simpson2021you, simpson2022tame} on social media platforms, including their experiences in navigating directly-experienced and perceived negative interactions with the platform itself. Karizat et al. \cite{karizat2021algorithmic} found users actively resisting the suppression of identity-based content on TikTok and that some users perceived harm in the overfitting of algorithmic content to their identities. DeVito \cite{devito2022transfeminine} describes these patterns of user experiences on TikTok through five theories, including the impact of \textit{algorithmic community} and \textit{identity flattening}, which describe how "TikTok’s algorithmic content distribution systems cluster users in a way that encourages the formation of communities" and "the reduction of one’s identity down to a limited set of socially acceptable attributes in environments where other attributes are discouraged or stigmatized to be accepted and treated well \cite{walker2020more}", respectively. Such experiences often lead to the practice of \textit{algorithmic resistance} \cite{velkova2021algorithmic}, where users seek to customize their platform experiences by actively "tricking" curation algorithms into subverting popular content it would otherwise recommend. Notably, much of the algorithmic resistance documented in literature reflects personal and/or collective resistance based on identity and ideology.

This study contributes three concrete extensions to the body of work surrounding theories of algorithmic resistance. First, we establish that poorly-fitted algorithmic content is indeed a problem for average-audience TikTok users. Second, we provide insights into users' tactical and behavioral methods for performing algorithmic resistance and pushing back against \textit{algorithmic persistence}. Third, we document the phenomenon of non-response from the TikTok algorithm. We describe algorithmic persistence in the context of theorization as the user-perceived persistence of the algorithm in delivering unwanted content even after users have put significant effort into resisting such content.

\subsection{Recommendations on the For You Page (FYP)}
TikTok’s driving force is the interest-driven FYP \cite{bhandari2022s,xu2019research}. TikTok’s core “hook” is the ability to view hand-picked content without much effort on the part of the viewer \cite{abidin2021mapping}. The exact mechanisms that drive the initial construction of the FYP have only recently been investigated by scholars. In the sign-up process, users are asked to choose from a set of genres of content and the FYP takes care of the rest \cite{zulli2022extending}. According to the platform \cite{tiktok_2020}, the contents of an individual's FYP are curated based on three factors: User interactions, Video information, and Device and Account settings. \citet{chen2022analysis} in their investigation of the algorithmic recommendation mechanisms on the FYP found that basic collaborative filtering, precise recommendations based on the users' social relations, and overlay recommendations based on the traffic pool were the core ways in which the FYP achieved efficient connection between the users and their interests. Overlay recommendations based on content traffic pool are perhaps one aspect that users of Facebook, Instagram, and Twitter might be less familiar with. This type of recommendation mechanism, which is based on the comprehensive weight of the content in terms of likes and comments, is “overlayed” into the FYP. When a creator posts a new video to TikTok, the video enters the first round of scoring by the algorithm, based on copywriting, topic, location, and more. If a video is “good” and the content is popular with users, the video will be pushed to more FYPs. This weighting is why some videos explode overnight and others do not \cite{chen2022analysis}. \citeyear{boeker2022empirical} experimented using a sock-puppet auditing approach to understand the factors that could potentially have an effect on the FYP, including language, current location, follow and like signals, as well as how the recommended content changes as a user watches or scrolls away, finding that who the users followed, what they liked, and video view rates had the strongest influence on FYP content. While TikTok affords common social features such as “liking” videos, following content creators, and sending direct messages, interpersonal connections are less prominent on the platform \cite{zulli2020rethinking}. 

Users have attempted to understand more about the FYP and how it may work. \citet{siles2022learning} analyzed algorithm awareness as a \textit{process} through a case study of the rollout of TikTok in Costa Rica. Many participants “expressed the sense that they had achieved [algorithmic] personalization relatively fast.” Some participants were “confused and surprised” when their expectations of the algorithm’s stability were not met. Interestingly, some participants expressed the desire to stop using TikTok altogether, citing the lack of new or different types of content \cite{siles2022learning}. \citet{issar2023social} explored the social construction of algorithms on TikTok and found that the algorithm was “seen as a way of being involved in either utilizing, mobilizing, or building TikTok communities." Additionally, it appeared as though users had several folk theories about the algorithm, many of which were competing against or contradicting each other. \citet{jones2023train} defined \textit{algorithmically imagined audiences}, combining theories about audience commodities with algorithmic imaginaries. Participants expressed theories about how to assert control over private algorithms and engage with community and counter-public world-making. An example of this may be a common phrase used across TikTok, “if you are seeing this, you are….” Typically, the sentence ends with “part of X community” or “support Y group.” The imagined audience is created through the creator’s understanding and theorizing about the way the TikTok algorithm works. 

\section{Methods}

\subsection{Participant Recruitment and Sampling}
We conducted recruitment with an interest survey on social media platforms such as TikTok, Reddit, Instagram, Twitter, Facebook, and various Slack groups accessible to the authors via university affiliation. In our recruitment survey, we asked participants to describe, in a short amount of text, specific types of content they consistently wanted to avoid and actively tried to avoid seeing in their TikTok recommendations. This open-ended question helped us identify participants who had reflected on their content preferences and could articulate specific experiences with unwanted content, ensuring rich and detailed interview data. Each participant self-identified their personal definition of "unwanted content" during the interviews, describing specific categories of content they actively tried to avoid on their For You Pages (FYPs). These ranged from content categories such as specific political viewpoints to particular lifestyle content they found irrelevant or distressing. Twenty-eight individuals indicated interest in being interviewed. We prioritized interviewing participants with the earliest availability as well as those that help us explore emerging theoretical concepts. Fourteen respondents were interviewed, as summarized in Table \ref{tab:participantsummary}.  Participants received US\$20 or equivalent in their preferred currency. The study was approved by our university's Institutional Review Board.

\subsection{Data Collection and Analysis}
We conducted semi-structured interviews remotely via Zoom between December 2022 and February 2023. Our interview protocol (Appendix \ref{app:protocol}) was iteratively refined based on emergent concepts during analysis. Using a systematic grounded analysis approach \cite{charmaz2012qualitative, corbin2014basics}, we conducted data collection and analysis concurrently. After each interview we performed line-by-line thematic coding, with each author coding interviews to ensure analytical rigor. We wrote analytical memos throughout the process to document our insights. As patterns emerged in the data we progressed to more focused thematic coding, refining our codebook throughout the process. After the fourteenth interview, we collectively determined that data saturation had been achieved. The emphasis on iterative data collection, interpretation, and analysis allowed us to continually refine our understanding of our participants' experiences and develop insights grounded in their perspectives.

Our analytical approach, inspired by grounded theory concepts, \cite{charmaz2012qualitative} was chosen for its effectiveness in exploring emergent phenomena. We focused on identifying patterns in participants' descriptions of unwanted content and their associated folk theories. Throughout the analysis, we employed systematic coding techniques and constant comparison, supported by memo-writing to ensure analytical depth and conceptual development.
\begin{table*}
    \centering
    \begin{tabular}{cccc}
         \textbf{Participant}&\textbf{Age Group}  &\textbf{Time Spent on TikTok per day}  &\textbf{Undesired Content} \\
         P1&18 to 24  &N/A (left TikTok)  &\#thatgirl and “perfect lifestyle“ topics \\
         P2&18 to 24  &30 to 60 minutes  &\#thirsttok and “adjacent“ topics \\
         P3&18 to 24  &60 to 90 minutes  &\#perfume, \#tattoo, and book spoilers \\
         P4&25 to 34  &Less than 15 minutes  &topics on Pakistani culture \\
         P5&18 to 24  &Over 120 minutes  &\#thirsttok and \#thirstrap \\
         P6&18 to 24  &15 to 30 minutes   &\#weightloss topics \\
         P7&25 to 34  &Over 120 minutes  &standup comedy \\
         P8&25 to 34  &Over 120 minutes  &\#antibiotics topics \\
         P9&25 to 34  &Over 120 minutes  &\#thirsttok \\
         P10&18 to 24  &60 to 90 minutes  &scary, horror, or thriller topics \\
         P11&25 to 34  &30 to 60 minutes  &shopping, shopping hauls, \#amazonfinds \\
         P12&25 to 34  &30 to 60 minutes  &too much of one type of video \\
         P13&25 to 34  &90 to 120 minutes   &\#dating, \#manifestation, \& \#marriage \\
         P14&25 to 34  &Less than 15 minutes  &\#conspiracy \\
    \end{tabular}
    \caption{Participant Summary}
    \label{tab:participantsummary}
\end{table*}

\section{Findings}

\subsection{Folk Theories in Unwanted Content Origins}
\subsubsection{Gravity of One-Time Searches}
One of the most prominent theories for why users saw a certain type of content was because, at one point, it was searched for using the TikTok search feature. This feature works like any other search engine, reading single words, phrases, or hashtags. Results are returned via a results page that allows the user to sort by "Top," "Videos," "Users," "Sounds," "Live," and "Hashtags." A user can click on a video result which opens a scrollable, results-only "feed" of videos. Some participants could directly trace their unwanted FYP content to searches on TikTok. As P8 mentioned:
\begin{quote}
    "My friend had my phone, and searched about [a health topic] happening to him. Since that day, the hashtag \#antibiotics [remained] on my page." (P8)
\end{quote}

However, participants who noticed this phenomenon were taken aback by how much weight a single search seemed to have on their FYPs. Users found that a single search for one distinct topic or hashtag could dominate their FYPs for weeks or months.
\begin{quote}
    "It feels kind of like they've over-emphasized that one search. It's not like you're searching for this all the time, or even really engaging with these videos [...] All of a sudden, for the next 5 weeks... they are curating your content for this particular thing." (P13)
\end{quote}

For P13 and other participants who felt this way, the surfacing of once-searched content from the distant past did not match the gravity of the action initially performed. While participants expected a search to perhaps surface similar content on the FYP for a short period of time, over a month of videos related to one search seemed excessive to them. 

\subsubsection{"Popular" Targeted Content}
Most participants understood that the TikTok algorithm might be surfacing topics or genres of videos that are "popular" at the moment and that TikTok may have an interest in promoting these types of viral videos. This is a basic principle of all recommender systems and was something our participants mentioned being familiar with. They also understood that viral content might be less tailored to their interests. As P11 mentioned: 
\begin{quote}
    "I see videos that are super viral right now.  You can just tell when one of these videos pops up immediately. They have a specific style." (P11)
\end{quote}

P11 could not theorize on any potential reason for why such videos were on their FYP, as they did not engage with them, beyond the idea that such videos were simply popular on TikTok at the time. Some participants noted a frequent, almost aggressive push from the algorithm to consume certain types of content. 

\begin{quote}
    "The thirst traps...really annoy me because I feel like it just assumes that this is just something that we want to watch as males, and it's just on my feed, whether I've looked it up or not, because I don't think anyone can possibly look it up." (P5)
\end{quote}

P5 added that they were especially uncomfortable receiving this type of content on their FYP as it is impossible to know if the subjects involved are adults. Participants mentioned \#thirsttok content coming up in different ways. P5 theorized that it was not just thirst trap content that the algorithm wanted them to conform to, but also tangentially-related content, such as videos from Jordan Peterson, a controversial psychologist and media personality.  
\begin{quote}
    "As for the male sigma videos, I think it's just generally a trend right now, which, as I said, I feel like I'm having to conform [to] because there's a lot of viewership for that kind of stuff like these Jordan Peterson videos that are totally taken out of context." (P5)
\end{quote}
\subsubsection{Unwanted Content By-Extension}
P5 and P6 believed that they might be receiving certain videos because their political ideology might align with \textit{opposition} to such content.  "Reaction" videos in the form of stitches, duets, and replies can often get packaged with the original, problematic content and thus surface ideologically relevant, but ultimately unwanted material, as P6 mentions. 

\begin{quote}
"I get content of men being like, "I'm not into women who are doing this" or just...lot of like really blatantly out of pocket, misogynistic comments that are happening, and it almost feels like a trend where people are trying to call out this behavior, but they're also simultaneously bringing attention to it." (P6)
\end{quote}
P11 describes a situation where they were served a set of videos that seemed to repackage short clips from popular television shows in the United States. P11 further added that friends had not sent them videos of this genre nor had they knowingly interacted with this kind of content. They concluded that this genre of TikTok was simply popular at the moment.
\begin{quote}
    "I think it’s just that the videos are super viral right now? I'm not sure. I usually scroll past them pretty quickly. You can tell when one of these videos pops up immediately. They have a specific style." (P11)
\end{quote}

Users also felt that the algorithm was pushing content that they \textit{should} be watching. P1 mentioned that their FYP was almost commanding them to adopt a certain lifestyle for themselves. 

\begin{quote}
    "I didn't really search for anything [other than my interests], but I started getting more of like \#thatgirl TikTok videos on it, whereas, like these girls... they'd wake up at 5 a.m. They like, drink whatever protein or green juice, smoothie, and they'd work out for an hour, and they'd be like med school students trying to study and figure out their day, or they'd be working corporate, and they'd work like PM tech jobs which are like super chill for a lot of the time." (P1)
\end{quote}

P1 recognized that \#thatgirl content might have also been related to other circles of content they were consuming such as tech job content. \#thatgirl content could not be pushed out of the FYP and resulted in P1 deleting their TikTok account.

\subsubsection{The Friend Effect}
Nearly all participants we spoke with reported that they maintained friendships through TikTok and mentioned receiving recommendations they believed were the results of videos exchanged, or videos their TikTok friends of friends' friends simply watched.  

\begin{quote}
    "I know a lot of my friends had also been looking at [unwanted content]. Since I had followed a few of [those friends], ... I got the other recommendations, and it's also what they're reviewing." (P1) 
\end{quote}

\textit{Collaborative filtering} describes  choosing to prioritize content that is either similar to previous content a user has viewed or to content others in a user's network have viewed \cite{herlocker2000explaining}. 

Participants also observed changes in their FYP because of new  connections and their interests, as P10 mentions.
\begin{quote}
    "I don't have a super high interest in cars to actually watch videos about it. But until I actually [got together]with my partner [...] I go to his page to look at his cars. And yeah, and soon after I think I got some [on my TikTok]...It's not a lot. But definitely started to curate for me." (P10)
\end{quote}

\subsubsection{Accidental Rewatching}
Because of the "looping" nature of the FYP, the same video can repeat multiple times due to user inattention \cite{price2022doomscrolling}. Our participants theorized that such accidental rewatching of some TikToks may have influenced some of the content on their FYP. 

\begin{quote}
    "Maybe you were watching something, and then you got distracted and left your phone and [TikToks] have that infinite loop where it just keeps playing and playing and playing. So it assumes I'm interested." (P13)
\end{quote}

P13 and a few others (P1 and P4) noted that accidentally rewatching videos may have caused the algorithm to incorrectly assume or overestimate their interest level in some types of content, and thus recommend more of the same genre.  

\subsection{Tactical Algorithmic Resistance}
Through our interviews, we found that participants utilized a range of tactics to remove unwanted content from their FYPs. Table 2 shows a quantitative breakdown of each tactic and how many participants in our sample used that tactic. Overall, scrolling away quickly was the most frequently used tactic for removing content, practiced by 9 out of 14 participants. Four participants described using positive reinforcement of desirable content such as commenting on categories of videos they wanted to see more of and searching for content they wished to add to their FYPs as a "replacement" for undesirable content. 
\begin{table*}[t]
    \centering
    \begin{tabular}{cc}
         \textbf{Tactics}& \textbf{No. of Participants} (out of 14)\\
         Scrolling away quickly& 9\\
         Positive reinforcement of desirable content& 4\\
         Reporting another user& 2\\
         Blocking another user& 2\\
         Filtering video keywords& 2\\
         Being intentional with follows& 2\\
         Reporting content& 2\\
         Using 'Not Interested' feature& 2\\
         Unfollowing problematic accounts& 2\\
 Contacting support&1\\
 Deleting the app&1\\
    \end{tabular}
    \caption{Tactics participants used to remove unwanted content}
\end{table*}

\subsubsection{Scrolling Away}
The most common strategy that participants used to indicate to TikTok that they did not want to see a certain type of content was to simply scroll away. Because they theorized that the duration of time spent on a video counted as a metric for the algorithm, all participants who used this approach emphasized the \textit{speed} of scrolling away.  
\begin{quote}
    "If there's any suggestion that I don't want to see, I just quickly scroll so TikTok doesn't get the idea that I even want to watch that. I swipe really fast." (P4)
\end{quote}

Scrolling away was perhaps the most popular strategy in this regard because of its embeddedness in the platform logic and ease relative to other tactics. Scrolling the FYP is generally how one receives (and pushes back upon) content.   

\subsubsection{Keyword Filtering}
P3 and P13 used an active approach to shaping their FYP out of repeated frustration with recommended content and a desire for more direct control. They used keyword filtering from the \textit{Content Preferences} tab in the TikTok app, where users can specify "filtered" video keywords for the content they no longer want to see, and any videos with that keyword would be removed from the FYP. 

P13 mentioned filtering out words like "marriage," "married," "engaged," "engagement," "wedding," "bride," and "bridesmaids" to indicate that they did not want to see anything in the genre of wedding videos. They mentioned coming up with the list themselves, not through careful observation of hashtags used in the videos they were receiving, but more so from their own idea of the kinds of keywords such videos would use. 

\begin{quote}
"I have a relationship. Marriage, married, engaged, engagement, wedding bride, bridesmaids group [blocked], because there was a period where all I was seeing was [this content], and I was like, I don't even know anybody who's trying to get married in time soon, so like, Why is this on my page?" (P13)
\end{quote}

Participants such as P3 and P13 expected that as a result of their filtering choices, the algorithm would infer the type of content they did not want to see, and avoid such recommendations. However, they both expressed that while the filtering function technically worked such that the set of specified keywords stopped appearing on their FYPs, they still saw content from the same genre, all with hashtag content that had not been blocked.

\subsubsection{Account Blocking}
Two participants mentioned using TikTok's blocking feature to rid themselves of certain content. Within TikTok, blocking a user or an account has the immediate effect of removing content from that account off one's FYP. No participants suggested that blocking specific accounts was an effective way of blocking content. P8 explained, in a conversation with their partner, that it is not the creators' fault that they appeared on his FYP. Creators cannot send content to a particular user's FYP. P8 went on to explain that blocking specific creators would be unfair; the creator did not ask to be displayed on a FYP that would not appreciate their content. 

\begin{quote}
"I really did want to block them. But you know I share this [account] with my wife, and my wife is also a creator. So she was like, `No, please don't do that. It's not their fault." (P8)
\end{quote}  

While blocking can be an effective strategy for managing specific creators, it is not a solution for blocking related content or accounts. As P7 mentioned, a specific creator kept appearing on his FYP even after taking action to block the account. After some time, he determined that the creator had established multiple accounts, which were all appearing on his FYP. 

\begin{quote}
"I had an issue with the person like I didn't like the person's contents, and you know the person kept on showing up on my [FYP]... I found out that the person's account was blocked. You know, it was! The account was taken down. And before I knew it, the person came up again...I was very surprised! [...] And, the person would actually come on [TikTok] to say. 'Yeah, my account was taken down, because...from a few people that reported my accounts." (P7)
\end{quote}

Other participants acknowledged that blocking specific creators would not solve their problem of removing genres of unwanted content. P10 mentioned that blocking did not affect the frequency of the undesirable content on her FYP. 

\begin{quote}
 "...every single time TikTok gave me [another undesirable video] it's from a different creator, but the under the same genre. Yeah. So you'd have to be blocking like, you know [all of them]." (P10)
\end{quote} 

\subsubsection{Account Reporting}
Either beyond or as opposed to blocking, some participants developed folk theories about leveraging TikTok's "report" feature as a mechanism to influence the algorithm's content selection. These theories centered on a belief that reporting would send a stronger signal to the system than other available feedback mechanisms. For instance, P12 mentioned following an individual who occasionally posted content they described as 'bigoted,' so they would report such videos based on their belief that this would train the algorithm to recognize and reduce similar content. 

P1 and P4 leveraged TikTok's ``Not Interested`` button on specific videos to indicate their displeasure with the recommendations, although they noted that the button was not easily discoverable and that they did not initially know where to find it when they first felt the need to use it. They considered this option a step before reporting. This suggests their folk theories about content control evolved as they discovered and experimented with different platform affordances 

\subsubsection{Unfollowing}
As a definite distinction between blocking and reporting, two participants mentioned unfollowing certain users who were associated with content that they no longer wanted to see. P9 implemented an unfollowing procedure that was quite rigorous. This participant noted when they began following certain creators and whether those had changed the content of their FYP.

Participants developed folk theories that aligned with the principles of collaborative filtering, a concept in which users with similar preferences receive similar content recommendations \cite{herlocker2004evaluating}. Their folk understanding suggested that following specific creators would associate their profile with certain content categories, and conversely, unfollowing would weaken these associations. However, all participants who spoke about unfollowing also demonstrated sophisticated folk theories about algorithmic temporality. They understood that it might take time for the algorithm to recognize and digest their unfollow actions, and thus changes to their FYPs might be slow to arrive.

\subsubsection{Non-Resistance: Positive Recommendations}
For some, the best approach towards removing unwanted content from their FYP seemed to be to indicate what types of content were \textit{wanted}. P3 and other participants intentionally re-watched videos they liked multiple times (as opposed to accidentally as mentioned above) to indicate to the algorithm that they would like to see more of such content. P4 went a step further by liking specific videos or following content creators who produced genres of content they enjoyed. For participants like P12, this also meant following creators from an entire side of TikTok, such as specifically following Black creators producing social/racial justice content, to inform the algorithm to serve more of such content.

\section{Discussion}
\subsection{Defining Algorithmic Persistence}

We observe individuals being unable to remove unwanted videos or genres of content from their FYPs, despite the active resistance to view such content, a phenomenon we refer to as \textit{algorithmic persistence}. We do not use "persistence" in the traditional sense, but rather as a label for the perceived \textit{unwillingness} of the TikTok algorithm to respond to user opposition to recommended content. Algorithmic persistence is not just a case of the algorithm missing optimization cues, but rather a phenomenon where structured user efforts fail to complete a feedback loop. 

This persistence is not simply an extension of popularity bias (\cite{abdollahpouri2019unfairness,zhang2021causal} because the nature of content that our users discussed wanting to remove from their FYPs was not restricted to only popular content. Rather, it demonstrates how the TikTok algorithm self-defines a content bubble for each user consisting of several genres of content and resists user attempts to escape it. From our interviews, it is also clear that a user seems to have some agency in the formation of the bubble. It is possible to draw loose cause-and-effect relationships between user actions and algorithmic recommendations, a finding also echoing prior research \cite{karizat2021algorithmic,milton2023see,simpson2022hey}. 
However, once captured in a content bubble, users found it difficult to leave, spending long periods of time being recommended undesired content. At best, algorithmic persistence is a minor annoyance. At worst, an app intended to bring joy, learning, and a sense of community can, through this unresponsiveness, exacerbate mental health conditions, trigger trauma responses, and force users to abandon the app altogether. 

\subsection{Algorithmic Persistence in the Context of Folk Theorization}

Our findings demonstrate how algorithmic persistence manifests and shapes users' folk theorization processes. While prior work has established that users develop folk theories to understand and adapt to algorithmic systems \cite{devito2018people, devito2021adaptive}, our results suggest that algorithmic persistence poses unique challenges to the theorization process. Specifically, we demonstrate how algorithmic persistence creates a situation where neither repeated folk theorization nor tactical strategies leveraging platform affordances for shaping one's TikTok experience lead to success in moderating one's FYP.

The interaction between algorithmic persistence and folk theorization thus appears to impact users' relationship with the TikTok platform. Past work has shown that users' willingness to adapt to algorithmic systems is, to some extent, influenced by their perception of the "platform spirit," or their understanding of what a platform is and what it is for \cite{devito2021adaptive}. Our findings suggest that algorithmic persistence may actively erode these perceptions over time, as some users come to see the FYP as deliberately resistant to their attempts at adaptation, as P1 described. The erosion of platform spirit may help explain why some users choose to limit their participation or leave platforms entirely.

These findings have important implications for how we think about algorithmic literacy and user adaptation. While previous work has emphasized helping users develop more sophisticated folk theories \cite{devito2021adaptive}, our findings also suggest that in the context of algorithmic persistence, theoretical sophistication alone may not be sufficient. We may need to consider how to help users develop what we might call "persistence-aware" folk theories that not only account for how one thinks a system might work, but also for their apparent resistance to tactics that would ordinarily indicate disinterest in a recommendation.

\section{Limitations and Future Work}
One limitation of our work is that given the scope of the study (14 interviews), the patterns of findings demonstrated above might not generalize widely across other algorithmically moderated content recommender systems. Furthermore, given how frequently the TikTok algorithm is perceived to change how it ranks and recommends content \cite{lee2022algorithmic}, it might be difficult to reproduce these results after a period of time. 

Furthermore, we present this theory of algorithmic persistence as an exploratory theme, generated through 14 user interviews. Such a sample size and methodology is not sufficient to make a definitive claim about an overarching phenomenon, and we encourage others to build upon this work. For instance, researchers could undertake a sock-and-puppet style audit similar to Boeker's experiment \citet{boeker2022empirical} to verify the presence of algorithmic persistence or similar phenomena. Future work may also extend this concept into other social media and shopping platforms.

\section{Conclusion}
In this paper, we sought to explore a phenomenon on TikTok where the algorithm serves up content to users' FYPs and, in the face of user resistance to such content, refuses to change its recommendation patterns. Through a combination of interviews with current and former TikTok users, we gathered evidence of this phenomenon being a common experience. We observed a range of practices being employed to mostly little to no success, and varying degrees of acceptance of sub-optimal FYPs to being forced off the app. We suggest that we may need to help users develop folk theories that account for how one believes a system might work and simultaneously how their disinterest tactics may seem ineffective. We encourage future work to understand the extent of this phenomena as well as its generalizability to other social platforms.

\begin{acks} 
We are grateful to our participants who were willing to share their personal experiences with us. Thank you to the members of the Communicative Practices in Virtual Workspaces Lab at the University of Washington for their support and encouragement: Soobin Cho, Anna Lindner, Joseph S. Schafer, Pitch Sinlapanuntakul, and Dr. Mark Zachry.  

This research was supported by a doctoral research grant from the University of Washington’s Department of Human Centered Design \& Engineering.
\end{acks}

\newpage

\appendix 
\section{Interview Protocol}\label{app:protocol}

\textbf{Demographics and Basics}
\begin{enumerate}
    \item How old are you?
    \item When did you first download it?
    \item What got you interested in downloading TikTok in the first place?
    \item About how many minutes or hours would you say you watch TikTok on a daily basis?
\end{enumerate}

\noindent
\textbf{Creatorship}
\begin{enumerate}
    \item Are you a creator?
    \item If so, have you ever considered becoming one?
\end{enumerate}
\noindent
\textbf{Friends and Social Connections}
\begin{enumerate}
    \item Are your friends also on TikTok?
    \item If so, do you communicate with them there often?
    \item If so, what kind of things do you share?
    \item What do those conversations with your friends look like? Can you describe the back-and-forth?
\end{enumerate}
\noindent
\textbf{For You Page}
\begin{enumerate}
    \item Generally, what kind of TikToks do you watch most often?
    \item Can you describe what other kinds of content you see on your FYP?
    \item Do you find the content to be pretty consistent day-to-day?
\end{enumerate}
\noindent
\textbf{Folk Theories}
\begin{enumerate}
    \item How do you think TikTok knows what content to serve you?
    \item Do you think the algorithm knows you better than you know yourself?
\end{enumerate}
\noindent
\textbf{Comparisons}
\begin{enumerate}
    \item Can you make some comparisons to Instagram/Instagram Reels/YouTube?
    \item What do you find to be significantly different or perhaps the same between these platforms?
\end{enumerate}
\noindent
\textbf{Actions Taken and Leaving \#Toks}
\begin{enumerate}
    \item Next, I want to talk about actually leaving the TikToks you no longer want to be on:
    \item What \#tok are you trying to leave?
    \item How would you describe the \#Tok you are trying to leave?
    \item How do you think you got there?
    \item Did you actively try to leave?
    \item How did you try to leave? What were some of the things you did to leave the \#tok?
    \item Is it hard to leave?
    \item Follow-up: Why do you think it’s so hard to leave?
\end{enumerate}
\noindent
\textbf{Results}
\begin{enumerate}
    \item Did you think you succeeded in leaving?
    \item Have these \#toks stuck with your FYP?
\end{enumerate}


\bibliographystyle{ACM-Reference-Format}
\bibliography{bibfile}

\end{document}